\documentstyle[11pt,aasms4]{article}

\begin{document}

\def\etal{et al. }
\def\astpart{{\it Astrop. Physics}}
\def\araa{{\it Ann.\ Rev.\ Astron.\ Ap.}}
\def\aplet{{\it Ap.\ Letters}}
\def\aj{{\it Astron.\ J.}}
\def\apj{ApJ}
\def\apjl{{\it ApJ\ Lett.}}
\def\apjs{{\it ApJ\ Suppl.}}
\def\aas{{\it Astron.\ Astrophys.\ Suppl.}}
\def\aa{{\it A\&A}}
\def\mnras{{\it MNRAS}}
\def\nature{{\it Nature}}
\def\pasa{{\it Proc.\ Astr.\ Soc.\ Aust.}}
\def\pasp{{\it P.\ A.\ S.\ P.}}
\def\pasj{{\it PASJ}}
\def\pre{{\it Preprint}}
\def\aph{{\it Astro-ph}}
\def\adspr{{\it Adv. Space. Res.}}
\def\expas{{\it Experimental Astron.}}
\def\ssr{{\it Space Sci. Rev.}}
\def\ar{{\it Astronomy Reports}}
\def\inpress{in press.}
\def\inprep{in preparation.}
\def\submit{submitted.}

\def\ap{$\approx$ }
\def\mjysr{MJy/sr}
\def\inu{{I_{\nu}}}
\def\inufit{I_{\nu fit}}
\def\fnu{{F_{\nu}}}
\def\bnu{{B_{\nu}}}
\def\msol{{M$_{\odot}$}}
\def\mic{{{\mu}m}}
\def\cm2{$cm^{-2}$}

\title{Archeops: a large sky coverage millimeter experiment for mapping Cosmic Microwave Background anisotropies}
\author{Xavier Dupac \altaffilmark{1}\\
Centre d'\'Etude Spatiale des Rayonnements (CESR)\\ 
9 av. du Colonel Roche, BP4346, F-31028 Toulouse cedex 4, France\\
\hspace{1em} \\
\& the Archeops collaboration}

\altaffiltext{1}{dupac@cesr.fr}

\begin{abstract}
We present Archeops, a balloon-borne bolometer experiment designed to map a large part of the sky at
millimeter and submillimeter wavelengths. 
The main scientific goal is to study the Cosmic Microwave Background anisotropies on all scales, in order to properly derive
the cosmological parameters of the Universe.
We present the sky coverage map of the first scientific flight.
\end{abstract}


\section{Introduction: the Cosmic Microwave Background radiation}
The Cosmic Microwave Background (CMB hereafter) is the oldest observable
electromagnetic radiation of our
Universe.
It was emitted about 300000 years after the Big Bang (z $\approx$
1000), when the Universe became transparent to photons.
The main cause for this radiation is the combination of hydrogen atoms at this
time from the proton and electron plasma.
Thus the light emitted was hot ($\approx$ 3000 K) but became gradually colder
with the expansion of the Universe, until being observed nowadays in the
millimeter domain (2.73 K).
This radiation was discovered in 1965 by Penzias \& Wilson in radiowaves.
The CMB is a very isotropic blackbody emission, but exhibits small fluctuations ($10^{-4}$
relative), whose large angular scales were observed by the COBE satellite
(Smoot \etal 1992).
The primordial quantum fluctuations were amplified by physical processes in
the cosmic soup at that time, such as acoustic wave phenomena.
These processes, and the way the information came to us today, are highly dependent on the cosmological
parameters of the Universe, so that the CMB map aspect and the CMB power
spectrum shape can provide powerful informations about the characteristics,
history and fate of our Universe.
These fluctuations are also the origin of the large-scale structure of the Universe (cluster and galaxy formation,
by gravitational collapse).

\section{The Archeops experiment}
Archeops is a balloon-borne experiment designed to fly in the
stratosphere to avoid the atmospheric contamination.
The telescope is two-mirror off-axis tilted Gregorian, with a primary
mirror providing an effective
aperture of 1.5 meters.
Twenty-two bolometers measure the millimeter and submillimeter radiation in
four frequency bands centered on 143 GHz (2.1 mm), 217 GHz (1.4 mm), 353 GHZ
(850 $\mic$) and 545 GHz (550 $\mic$).
These detectors are installed on the focal plane stage cooled at 0.1 K by
helium dilution techniques.
The instrumental setup and the calibration are detailed in Beno\^\i t \etal (2001).

Several recent balloon-borne experiments have concentrated on the small
angular scales of the CMB, such as BOOMERANG (De Bernardis \etal 2000) and
MAXIMA (Hanany \etal 2000).
The main scientific goal of Archeops is rather to investigate a broad band of
angular scales, in order to make the link between the COBE full-sky survey and
the small-scales experiments of today and the near future.
For this the aim is to map a large part of the sky ($\approx$
30 \%) with unprecedented angular resolution (5'-10').
To achieve this goal, the scanning strategy is as follows: the gondola rotates
at 2 rpm, allowing the beam to make large circles on the sky.
These circles slowly shift with respect to the sky, because of the rotation of
the Earth and the moving of the balloon on the Earth.
Because of the high sensitivity needed, the experiment has to fly during the
night.
Since we want as long flights as possible in order to increase the total
integration time, the polar night is the best.
That is why the Esrange base in Kiruna (Swedish Lappland) has been chosen for
launching Archeops.
However, the first (technological) flight was done from Trapani (Sicily) in
July 1999, and gave 4 hours of data.
The first scientific flight was done from Kiruna in January 2001, and gave 8
hours of data.
Longer flights are possible with normal atmospheric conditions, and expected
for the winter 01/02.

\section{Sky coverage of the Kiruna flight}
We present in Fig. 1 a simulated map at 2 mm (143 GHz) containing the CMB (including the dipole
due to the Earth movement towards the cosmic fluid) and the Galaxy, which are
the two major components of the millimeter sky.
On this map, we have overimpressed the sky coverage of the Kiruna scientific flight.
The projection is Galactic Mollweide centered on the Galactic anticenter.
This map shows that a large part of the CMB fluctuations sky has been observed
by Archeops.
The data are still being analyzed.

\begin{figure}[ht]
\epsfbox{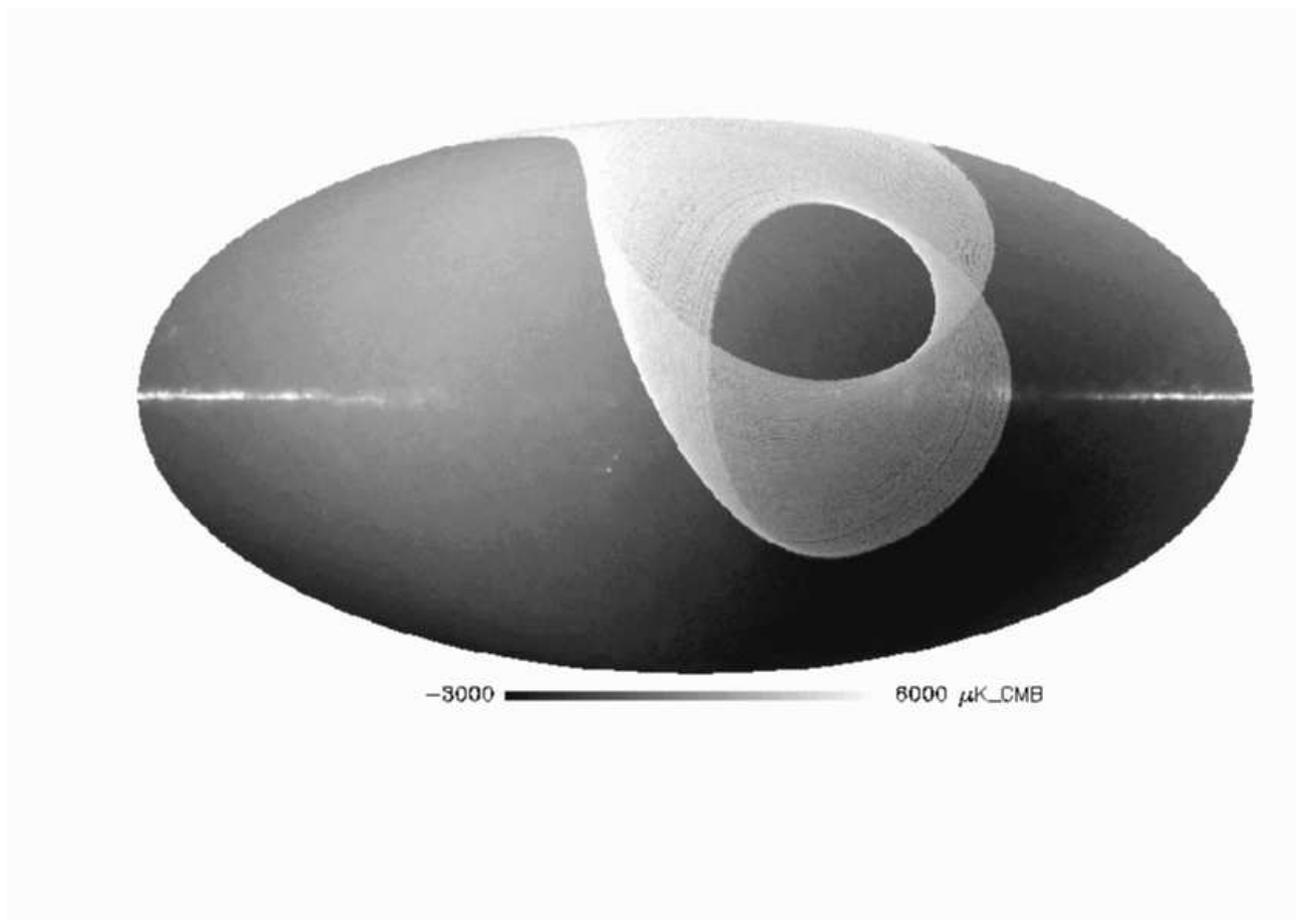}
\caption[]{Simulated map at 2 mm (143 GHz) containing the CMB (dipole+fluctuations) and the Galaxy.
We have overimpressed the sky coverage of the Kiruna scientific flight.
The projection is Galactic Mollweide centered on the Galactic anticenter.
}
\end{figure}

\end{document}